\documentclass{JHEP3}

\usepackage{amssymb}
\usepackage{amsfonts}
\usepackage{amsbsy}
\usepackage{amsmath}
\usepackage{amsthm}
\usepackage{graphicx}
\usepackage{epstopdf}
\usepackage{multirow}
\usepackage{latexsym}
\usepackage{array}
\input cyracc.def %sha
\newfont{\twelvecyr}{wncyr10 at 12pt}
\def\sha{\text{\twelvecyr\cyracc{Sh}}}

\def\Z{\mathbb{Z}}
\def\F{\mathbb{F}}
\def\Q{\mathbb{Q}}

\def\G{\mathfrak{g}}

\def\P{\mathbb{P}}

\def\n3a{t}

\def\ge{{\mathfrak{e}}}
\def\gso{{\mathfrak{so}}}
\def\gsu{{\mathfrak{su}}}
\def\gsp{{\mathfrak{sp}}}
\def\gf{{\mathfrak{f}}}
\def\gg{{\mathfrak{g}}}

\newcommand{\eq}[1]{(\ref{#1})}

\title{Classifying bases for 6D F-theory models}

\author{David R.  Morrison$^{1}$ and Washington Taylor$^2$\\
$^1$Departments of Mathematics and  
Physics\\ University of California, Santa Barbara\\ Santa Barbara, CA 93106, USA\\
\\
$^2$Center for Theoretical Physics\\
Department of Physics\\
Massachusetts Institute of Technology\\
77 Massachusetts Avenue\\
Cambridge, MA 02139, USA\\
\\
{\tt drm} {\rm at} {\tt math.ucsb.edu},
{\tt wati} {\rm at} {\tt mit.edu}
}

\preprint{UCSB Math 2012-01, MIT-CTP-4339}

\abstract{We classify six-dimensional F-theory compactifications in
  terms of simple features of the divisor structure of the base
  surface of the elliptic fibration.  This structure controls the
  minimal spectrum of the theory.  We determine all irreducible
  configurations of divisors (``clusters'')
that are required to carry nonabelian
  gauge group factors based on the intersections of the divisors with
  one another and with the canonical class of the base.  All 6D
  F-theory models are built from combinations of these irreducible
  configurations.  Physically, this geometric structure characterizes
  the gauge algebra and matter that can remain in a 6D theory after
  maximal Higgsing.  These results suggest that all 6D supergravity
  theories realized in F-theory have a maximally Higgsed phase in
  which the gauge  algebra is built out  of summands of the
  types $\gsu(3), \gso(8), \gf_4, \ge_6, \ge_8, \ge_7, (\gg_2 \oplus
  \gsu(2)),$ and $\gsu(2) \oplus \gso(7) \oplus \gsu(2)$, with minimal
  matter content charged only under the last three types of summands,
  corresponding to the non-Higgsable cluster types identified through
  F-theory geometry.  Although we have identified all such geometric
  clusters, we have not proven that there cannot be an obstruction to
  Higgsing to the minimal gauge and matter configuration for any
  possible F-theory model.  We also identify bounds on the number of
  tensor fields allowed in a theory with any fixed gauge algebra; we use
  this to bound the size of the gauge group (or algebra) in a simple class of
  F-theory bases.}

\begin{document}

%--------------------------------
\section{Introduction}

In a series of recent works, progress has been made in systematically
analyzing the space of possible consistent six-dimensional ${\cal N} =
1$ supergravity theories \cite{Grassi-Morrison}-\cite{Grimm-6D}.  A key
part of this work has been the close correspondence relating the
spectrum and Green--Schwarz couplings of 6D supergravity theories to
geometric features of F-theory constructions.

F-theory \cite{Vafa-f, Morrison-Vafa, Morrison-Vafa-II} is a very
powerful, nonperturbative approach to constructing string vacua in
even-dimensional space-time.  It is known that there is a finite set
of possible combinations of gauge groups and matter content that can
arise in F-theory constructions of 6D supergravity theories
\cite{Grassi91,Gross, KMT-II}.  Nonetheless, there is no systematic global
characterization of the space of such theories.  6D F-theory
constructions are based on a choice of elliptically fibered Calabi--Yau
manifold over a base $B$ that is a complex surface.  In general, for
any given base $B$, there is an enormous range of possible gauge
groups and matter content that can be realized in the corresponding 6D
supergravity theory.  The massless gauge group and matter content of
the gravity theory can be changed by giving expectation values to
charged matter fields; this reduces the gauge symmetry and massless
matter content through the familiar ``Higgsing'' process.  In the
F-theory picture, models with different spectra over the same F-theory
base can be arranged by tuning the coefficients in a Weierstrass
description of the model to arrange for certain codimension one and
codimension two singularities in the elliptic fibration.  Relaxing the
constraints on coefficients needed for larger gauge symmetry
corresponds to Higgsing in the 6D supergravity theory.  Several recent
papers \cite{0, Morrison-Taylor, Braun} explore the range of gauge
groups and matter content available for F-theory models that are
realized over the simplest base surface, $B = \P^2$.

Just as 6D theories with different spectra are connected by Higgsing
transitions, more exotic transitions involving tensionless strings
connect the F-theory models on different base manifolds \cite{dmw,
  Seiberg-Witten, Morrison-Vafa-II}.  In this paper we carry out a
systematic analysis of the kinds of F-theory models that can arise by
restricting attention to the most generic theory over each base
manifold $B$.  Over any given base $B$, there is a unique gauge algebra
and matter representation for the generic theory, characterized
physically by Higgsing (i.e., giving a vacuum expectation value to) all
possible matter fields.  In some cases, maximal Higgsing completely
removes any gauge symmetry.  In other cases, maximal Higgsing removes
all charged matter fields from the massless spectrum.  And in yet
other cases, some small amount of
residual charged matter fields remain that cannot be
lifted by Higgsing.  We use the structure of effective divisor classes
in $B$ with negative self-intersection to determine a minimal gauge
algebra and matter content that must be present for any F-theory model
over each base $B$.  By thus identifying the geometrical structures
underlying all possible ``non-Higgsable clusters''
composed of gauge
group factors and matter fields that cannot be Higgsed or factorized,
we develop tools for analyzing the space of all F-theory
compactifications.

6D supergravity theories are characterized by the number $T$ of tensor
multiplets in the  theory.  Theories with many tensor multiplets can
satisfy anomaly cancellation through a generalization of the
Green-Schwarz mechanism \cite{Sagnotti, Sadov}.
The number of tensor multiplets in a 6D supergravity theory coming
from an F-theory compactification is related to the topology of the
base $B$ through
\begin{equation}
T = h^{1, 1} (B) -1 \,.
% % \label{eq:}
\end{equation}
As the number of tensor multiplets increases, the topological
complexity of the F-theory base $B$ grows, and the generic (completely
Higgsed) models become more complicated.  We show here that for any
given gauge algebra $\mathfrak{g}$ and matter content that can be realized in a
maximally Higgsed theory, there is a bound on the value of $T$ for any
F-theory construction with this spectrum.  
Combining this result with knowledge of the set of possible non-Higgsable
clusters (``NHC's'') suggests a path to systematic classification of
all 6D F-theory models.
While the finiteness of the
set of F-theory models indicates that there is a maximum value
possible for $T$ compatible with a valid F-theory construction, no
specific upper bound has previously been determined for this parameter.  The
maximum value ever found for a consistent F-theory construction or any
other quantum 6D supergravity construction is $T = 193$ \cite{Candelas-pr,
  Aspinwall-Morrison-instantons}.  
We show here that the bounds determined on  $T$ from the gauge group limit
the size of a class of related models constructed from a linear chain
of divisors.
In a sequel to this paper \cite{mt-toric}, we will
systematically analyze and enumerate all
F-theory models over toric bases, and show that  the model with
$T = 193$  fits naturally into this classification.

Section \ref{sec:preliminaries} reviews some basic aspects of F-theory
and algebraic geometry that are needed for the analysis in this paper.
In Section \ref{sec:general} we analyze the intersection properties of
divisors on the base $B$ and determine all irreducible clusters for
maximally Higgsed theories and the associated gauge algebra and matter
contents.  In Section \ref{sec:bounds}, we determine upper bounds on
$T$ for any given gauge algebra.  Section \ref{sec:conclusions} contains
some concluding remarks.

\section{Geometric and F-theory preliminaries}
\label{sec:preliminaries}

%\subsection{F-theory compactifications to six dimensions}

Pedagogical introductions to the aspects of F-theory needed for this
paper can be found in \cite{Morrison-TASI, Denef-F-theory, WT-TASI}.
We review here briefly a few of the most salient points.

An F-theory compactification to six dimensions is defined by an
elliptically fibered Calabi--Yau threefold with section over a base
$B$.  The fibration can be described by a Weierstrass equation
\begin{equation}
y^2 = x^3 + fx + g \,
\label{eq:Weierstrass}
\end{equation}
where $f, g$ are local functions on a complex surface forming the base $B$.  
Globally, $f, g,$ and the discriminant locus
\begin{equation}
\Delta = 4 f^3 + 27 g^2
\label{eq:discriminant}
\end{equation}
are sections of line bundles
\begin{equation}
f \in -4K, \; \; g \in -6K, \; \; \Delta \in -12K
% % \label{eq:}
\end{equation}
where $K$ is the canonical class of $B$.
The  canonical class $K$ satisfies
\begin{equation}
K \cdot K = 9-T
% % \label{eq:}
\end{equation}
where the inner product is the intersection form on $H^2 (B,\Z)$ and
$T$ is the number of tensor multiplets in the 6D supergravity theory.

The gauge
group in F-theory is a compact reductive group $G$ whose component group
$\pi_0(G)$ coincides with the Tate--Shafarevich group $\sha_{X/B}$
of the fibration \cite{triples}, whose fundamental group
$\pi_1(G)$ coincides with the Mordell--Weil group $\operatorname{MW}(X/B)$
of the fibration \cite{pioneG}, and whose Lie algebra $\mathfrak{g}$
has a non-abelian part that is
 determined by the singular fibers in codimension $1$ as
described below.  To avoid the subtle analysis involving 
 the Tate--Shafarevich and Mordell--Weil groups
that is necessary for the full specification
of the gauge group, we focus in this paper on the nonabelian gauge
algebra of the F-theory model.

Along the discriminant locus $\Delta$ the elliptic fibration is
singular.  Codimension one singularities carry nonabelian gauge algebra
summands of the 6D theory, and codimension two singularities carry
matter fields.  The gauge algebra along a given component $C$ of the
discriminant locus can be determined by the Kodaira classification and
the Tate algorithm \cite{Kodaira, Morrison-Vafa-II, Bershadsky-all,
Morrison-sn, Grassi-Morrison-2} in terms of the degrees of vanishing
of $f, g, \Delta$ along the curve $C$ (see Table~\ref{t:Kodaira}).  In
some cases further information regarding monodromy is needed to
determine the precise gauge algebra; monodromies can give rise to
non-simply-laced algebras in some situations \cite{Bershadsky-all,
Grassi-Morrison-2}.  The possible singularity types at codimension two
are not completely classified.  In most simple cases, a local rank one
enhancement of the gauge algebra gives matter that can be simply
interpreted \cite{Bershadsky-all, Katz-Vafa, Grassi-Morrison}, but in
other cases the singularities can be more complicated.  Recent
progress in understanding codimension two singularities and associated
matter content appears in \cite{Morrison-Taylor, Esole-Yau,
Grassi-Morrison-2, Esole-Yau-II}.

\begin{table}
\begin{center}
\begin{tabular}{|c |c |c |c |c |}
\hline
ord ($f$) &
ord ($g$) &
ord ($\Delta$) &
singularity & nonabelian symmetry algebra\\ \hline \hline
$\geq $ 0 & $\geq $ 0 & 0 & none & none \\
0 & 0 & $n \geq 2$ & $A_{n-1}$ & $\gsu(n)$  or $\gsp(\lfloor n/2\rfloor)$\\
 $\geq 1$ & 1 & 2 & none & none \\
1 & $\geq 2$ &3 & $A_1$ & $\gsu(2)$ \\
 $\geq 2$ & 2 & 4 & $A_2$ & $\gsu(3)$  or $\gsu(2)$\\
$\geq 2$ & $\geq 3$ & $6$ &$D_{4}$ & $\gso(8)$ or $\gso(7)$ or $\gg_2$ \\
2 & 3 & $n \geq 7$ & $D_{n -2}$ & $\gso(2n-4)$  or $\gso(2n -5)$ \\
 $\geq 3$ & 4 & 8 & $\ge_6$ & $\ge_6$  or $\gf_4$\\
3 & $\geq 5$ & 9 & $\ge_7$ & $\ge_7$ \\
 $\geq 4$ & 5 & 10 & $\ge_8$ & $\ge_8$ \\
\hline
$\geq 4$ & $\geq6$ & $\geq12$ & \multicolumn{2}{c|}{ does not occur in 
F-theory } \\ 
\hline
\end{tabular}
\end{center}
\caption[x]{\footnotesize  Table of singularity types for elliptic
  surfaces and associated nonabelian symmetry algebras.}
\label{t:Kodaira}
\end{table}

The components $C$  of the discriminant locus
carrying nonabelian gauge algebra summands in an
F-theory model are irreducible effective divisors in $B$.  The
discriminant locus itself, $\Delta = -12K$, is effective but need not
be irreducible.
The key feature of the algebraic geometry of surfaces that will  be
useful to us here relates to irreducible effective divisors of $B$.
If $C$ is an irreducible effective divisor of $B$ satisfying $C \cdot
C < 0$, and $A$ is an effective divisor satisfying $A \cdot C < 0$,
then $C$ is an irreducible component of $A$, meaning that
\begin{equation}
C \cdot C < 0, \;\;\; A \cdot C < 0 \; \; \; \Rightarrow \; \; \;
A = C + X
% % \label{eq:}
\end{equation}
with $X$ effective.  We will use this fact repeatedly in our analysis,
generally using it to show that certain divisors must be contained in
$-4K, -6K,$ and $-12K$ and thus carry a minimal gauge algebra.  For
example, consider an irreducible effective divisor $C$ satisfying $C
\cdot C = -8$.  The genus of $C$ is fixed by the relation
\begin{equation}
(K + C) \cdot C = 2g-2.
 \label{eq:genus}
\end{equation}
If $C$ is a rational curve (topologically $\P^1$, with $g = 0$) having
$C \cdot C = -8$, then $K \cdot C = 6$.  It follows that $-4K \cdot C
= -24$, so that $-4K = 3C + X_4$ with $X_4$ effective and $X_4 \cdot C
\geq 0$.  Similarly, $-6K = 5C + X_6$ and $-12K = 9C +X_{12}$.  Thus,
$f, g,$ and $\Delta$ have degrees of vanishing at least $3, 5, 9$ on
$C$, so $C$ carries an $\ge_7$ gauge algebra.  A similar argument shows
that any irreducible effective divisor $C$ with $C \cdot C < -2$ must
carry a nonabelian gauge algebra summand.  This fact is mentioned in a
related physics context in \cite{Cordova}.

A particularly simple set of F-theory bases are given by complex
projective space $\P^2$ and the Hirzebruch surfaces $\F_m$.  $\P^2$ is
the only F-theory base with $T = 0$, and has a cohomology ring
generated by a single divisor $H$ with $H \cdot H = 1$.  The
Hirzebruch surface $\F_m$ is a $\P^1$ fibration over $\P^1$ that
contains an effective irreducible divisor $D$ with $D \cdot D = -m$
corresponding to a section of the fibration.  Each Hirzebruch surface
also contains an effective irreducible divisor $F$ corresponding to a
fiber, satisfying $D \cdot F = 1, F \cdot F = 0$.  Together $D$ and
$F$ span the second cohomology of $\F_m$.  The Hirzebruch surfaces
with $m \leq 12$ are the only F-theory bases with $T = 1$.  All
F-theory bases for 6D theories (except the Enriques surface, which
supports models with no gauge group or matter content) can be found by
blowing up a sequence of points on one of the bases just described.
When the degrees of vanishing of $f, g, \Delta$ exceed $4, 6, 12$ at a
point, the singularity of the elliptic fibration over $B$ is so bad
that any corresponding F-theory model would have tensionless strings.
To get a good F-theory model from such a fibration, one must
blow up the point in the base,
increasing the number of tensor multiplets by one, and giving a ``$(-1)$-curve'' (rational curve with $C \cdot C = -1$) on which the degrees of
vanishing of $f, g, \Delta$ are reduced by $4, 6, 12$ from the
original singular point.\footnote{The tensionless
string transitions \cite{dmw,
  Seiberg-Witten, Morrison-Vafa-II} connect two components of F-theory
moduli space: on one of these, we allow more general polynomials
$f$ and $g$ to avoid the highly singular point; on the other component,
we blow up, increasing the number of tensors in the spectrum.  The transition
point itself is not considered to be an F-theory model.}
  When the degrees of vanishing of $f, g,
\Delta$ exceed $4, 6, 12$ on a curve, the singularity is even worse,
and the remedy is to divide the coefficients $f$ and $g$ in the Weierstrass
equation by appropriate powers of the equation of the curve.  Unfortunately,
that change alters the canonical bundle of the total space, and so leaves
one with an elliptic fibration whose total space is not Calabi--Yau, so that
it cannot be used in F-theory at all.  This is indicated in the final line
of Table~\ref{t:Kodaira}.

\section{Non-Higgsable clusters}
\label{sec:general}

On any F-theory base, there is a moduli space of theories in which the
generic model has all possible matter fields Higgsed.  This does not
necessarily mean that the generic model has no matter fields, only
that no further Higgsing is possible in the generic configuration.
Information about the maximally Higgsed model on a given base is
contained in the intersection structure of effective divisors on the
base.  On any base, an important feature of the geometry is the set of
irreducible effective curves with negative self-intersection.  As
noted in the previous section, any irreducible effective divisor $C$
with $C \cdot C < -2$ must carry a nonabelian gauge algebra summand.  In
this section we consider all possible intersecting combinations of
irreducible effective divisors with $C \cdot C < -1$ that can arise on
valid F-theory bases and the minimal gauge algebras and matter content
associated with these clusters.  This gives a set of ``non-Higgsable
clusters'' of gauge algebras and matter content that can appear as
factors in maximally Higgsed 6D supergravity theories.

\subsection{Clusters of intersecting irreducible divisors}

Each specific configuration of irreducible effective curves $C_i$,
characterized by the self- and pairwise intersection numbers of the
curves $C_i \cdot C_j$, gives rise to a specific minimal gauge and
matter content that can be computed by determining the multiplicity
with which each $C_i$ appears in the multiples $-4K, -6K,$ and $-12K$
of the canonical class.  These gauge algebras and matter content appear
in the generic non-Higgsable theory on the given base.  Note that for
any set of curves with all self-intersections equal or greater than
-2, $-nK \cdot C_i \geq 0$, and none of the curves $C_i$ need to occur
as components of $-nK$, so there is no nonabelian gauge algebra required
on such a set of curves.  Thus, we focus attention here on clusters of
intersecting curves that include at least one curve of
self-intersection $-3$ or below.  In the following analysis all curves
mentioned are irreducible and effective unless otherwise stated
(except the anti-canonical divisor $-K,$ and residual components denoted
$X, Y$, which are effective but not irreducible).

\subsubsection{Single irreducible divisors}

As discussed above, a single irreducible effective curve $C$ with $C
\cdot C < -2$ must have sufficiently high degrees of vanishing of $f,
g, \Delta$ to support a contribution to the nonabelian gauge algebra.  
If such a
curve $C$ does not intersect other curves with negative
self-intersection, then the degrees of vanishing and the associated
gauge algebra are easily computed.
The results of this computation are
familiar from the F-theory bases $\F_m$, which give theories dual to
heterotic compactifications having known generic gauge algebra and
matter configurations \cite{Kachru-Vafa, Morrison-Vafa-II}.  Each base
$\F_m$ has a single irreducible effective divisor $D$ with negative
self-intersection $D \cdot D = -m$; the calculation of the
multiplicities with which $D$ appears in $f, g, \Delta$ follows just
as in the example $C \cdot C = -8$ below Equation (\ref{eq:genus}).
The gauge algebra and matter content for isolated curves with $C \cdot C
= -m, 3 \leq m \leq 12$ are tabulated in Table~\ref{f:primitives}.
(Note that in some cases the degree of $\Delta$ is increased by the
fact that deg $\Delta \geq$ Min (3 deg $f$, 2 deg $g$).)

A simple way to determine the multiplicities of the curve $C$ in $-nK$
is to write $ -K$ as
\begin{equation}
-K = \gamma C + Y \,,
% \label{eq:}
\end{equation}
where $Y \cdot C = 0$ and $\gamma$ is taken over the rational
numbers.  From $-K \cdot C = 2-m, C \cdot C = -m$, it follows that
$\gamma = (m-2)/m$.  Writing $-nK$ as an integral multiple of $C$ plus
a residual part $X$ that satisfies $X \cdot C \geq 0$
\begin{equation}
-nK = cC + X \,,
% \label{eq:}
\end{equation}
it follows that $c =\lceil n (m-2)/m\rceil$.  Thus, the degrees of
vanishing of $-4K, -6K,$  and $-12K$ over an isolated curve of
self-intersection $-m$ are
\begin{equation}
[f]=\lceil  4 (m-2)/m\rceil, \;\;\;
[g]=\lceil 6(m-2)/m\rceil,\;\;\;
[\Delta]=\lceil 12 (m-2)/m\rceil \,.
% \label{eq:}
\end{equation}
Note that for $m = 9, 10, 11$ the degrees of vanishing are $4, 5, 10$,
corresponding to an $\ge_8$ singularity on $C$, but the residual divisor
$X$ must have nonvanishing intersection with C; this raises the degree
of vanishing at the intersection point to $4, 6, 12$ so that the
intersection point must be blown up; thus the only value of $m > 8$
possible in a good F-theory base is $m = 12$.  Physically this
corresponds to the fact that there is no fundamental matter field for
$\ge_8$ that would arise from the intersection of an $m = 9, 10,$ or
$11$ curve with another component of the discriminant locus.
The derivation of the gauge algebra for individual curves with
self-intersection $-m$ is described in
section~\ref{sec:spectra}.  

Note also that effective irreducible curves of negative self-intersection
appearing in F-theory bases must be rational; higher genus curves
automatically carry degrees of vanishing of $f, g,$ and $\Delta$ that
are equal to or greater than $(4, 6, 12)$, as can be verified by a
simple computation.
Assume that there is a curve $C$ with $C \cdot C < 0, g > 0$.  Then
from \eq{eq:genus} it follows that $K \cdot C  \geq -C \cdot C$.  The
effective divisor $-nK$ must then contain the component $C$ at least
$n$ times, since if $-nK = cC +  X$, we have $X \cdot C = -nK \cdot
C-cC \cdot C < 0$ when $C \cdot C < 0$ unless $c \geq n$.

\subsubsection{Pairs of intersecting divisors}
\label{sec:pairs}

In addition to isolated curves, 
there are combinations of
intersecting curves that have higher degrees of vanishing for $f, g,
\Delta$ than are required by the self-intersections of the individual
curves.  
We next consider pairs of curves $A, B$ that each have negative
self-intersection and that intersect one another in at least one point
\begin{eqnarray}
A \cdot A & = &  -x < 0 \label{eq:aa}\\
B \cdot B & = &  -y < 0\\
A \cdot B & = &  p > 0 \label{eq:ab}
\end{eqnarray}

For example, consider the case of two curves, each with
self-intersection $-3$, that intersect at a single point ($x = y = 3,
p = 1$).  In this case, writing for example $-4K = aA + bB + X$, with
$X$ effective and $X \cdot A \geq 0, X \cdot B \geq 0$ so that $X$ does not
contain $A$ or $B$, we have from \eq{eq:genus} that $-4K \cdot A = -4K
\cdot B= -4$, from which it follows that $3a-b \geq 4, 3b-a \geq 4$
and thence that $a + b \geq 4$, so that $f$ is of degree at least four
on the intersection point $A \cdot B$.  Similarly $g, \Delta$ have
degrees at least 6 and 12, so the intersection point carries an
elliptic fibration structure that is too singular and the base must be
blown up for a valid F-theory construction.  There may be a model on
the blown up base with a higher value of $T$ that is consistent, but
this shows that no consistent F-theory model will have a base
containing two intersecting curves of self-intersection $-3$.

It is straightforward to confirm that making either curve have a more
negative self-intersection or increasing the number of intersection
points between the curves simply makes the model more singular.
Taking the general case described by (\ref{eq:aa}-\ref{eq:ab}), and
writing again
\begin{equation}
-nK = aA + bB + X
% \label{eq:}
\end{equation}
we have
\begin{eqnarray}
-nK \cdot A & = & n (2-x) = -ax  +bp + X \cdot A,\\
-nK \cdot  B & = & n (2-y) = ap   - by + X \cdot B \,.
\end{eqnarray}
This leads to a pair of inequalities
\begin{eqnarray}
ax-pb & \geq & n (x-2), \\
by-pa & \geq &  n (y-2) \,.
\end{eqnarray}
This gives lower bounds on $a, b$ when $p^2 < xy$
\begin{eqnarray}
a & \geq &  \frac{n}{xy-p^2}  (xy + py-2y-2p) \,,\\
b & \geq &  \frac{n}{xy-p^2}  (xy + px-2x-2p) \,, \label{eq:ab-bounds}
\end{eqnarray}
and no  solution if $p^2 > xy$. 
There are three marginal solutions where $p^2 = xy$;
in the  cases
$x = y = p = 1$ of a ($-1$)-curve intersecting a ($-1$)-curve at a single
point, and $x = y = p = 2$ of two ($-2$)-curves intersecting at two
points the solution $a = b = 0$ gives a valid configuration,
and when $x = 4, y = 1, p = 2$ there is a solution with $b = 0$ and
$a$  as for an  isolated $(-4)$-curve.
In the cases where $p^2 < xy$,
the degree of $-nK$ at the intersection point $A \cdot B$ is
given by
\begin{equation}
a + b \geq \frac{n}{xy-p^2}  (2xy + p (x + y)-4p-2x-2y) \,.
% \label{eq:}
\end{equation}
For a valid F-theory base
we need $a + b < n$ for at least
one of $n = 4, 6, 12$, which is only possible
if
\begin{equation}
(x + p-2) (y + p-2) < 4 \,.
% \label{eq:}
\end{equation}
Thus, the only possible pairs of negative self-intersection curves
that can arise with a single intersection ($p =1$) in an F-theory base
are those with self-intersections $(-x, -y) = (-3, -2), (-2, -2),$ or
$(-m, -1)$ with $m \leq 12$.  Double intersections ($p = 2$) are only
possible between curves with self-intersections $(-x, -y) = (-m, -1)$
with $m = 1, 2, 3$ or 4.  Triple intersections are not possible.  It
is easy to check that in all cases containing a ($-1$)-curve, the
$(-1)$-curve need not be a component of $-nK$, so that the degrees of
$ f, g, \Delta$ and the nonabelian gauge algebra on the other component
with self-intersection $-m$ are the same as if that curve were
isolated.

The only new irreducible cluster that has thus arisen containing a
pair of intersecting negative self-intersection curves where a
nonabelian gauge factor must arise is the intersection of a ($-3$)- and ($-2$)-curve.  The degrees of $f, g, \Delta$ on the two curves in this
configuration can be computed using the minimum values of $a, b$
satisfying \eq{eq:ab-bounds}.  Computation of the gauge algebra and
matter content on this intersecting divisor cluster requires
consideration of the monodromy structure; the details of this analysis
are described in Section \ref{sec:spectra} below.  The result is that
the intersecting $(-3)$- and $(-2)$-curves carry a gauge algebra $\gg_2 \oplus
\gsu(2)$, with $16$ half-hypermultiplet matter fields transforming under
the ${\bf 7} + {\bf 1}$ of $\gg_2$ and the fundamental of $\gsu(2)$.  This
configuration and the associated gauge algebra is depicted in
Figure~\ref{f:clusters}, and listed in Table~\ref{f:primitives}.

\subsubsection{More than two intersecting divisors}

\begin{figure}
\begin{center}
\begin{picture}(200,110)(- 93,- 55)
%\grid
\thicklines
\put(-175, 25){\line(1,0){50}}
\put(-150,32){\makebox(0,0){$-m$}}
%\put(-150, 18){\makebox(0,0){$(m = 3, 4, 5, 6, 7, 8, 12)$}}
\put(-150,-33){\makebox(0,0){\small $\gsu(3), \gso(8), \gf_4$}}
\put(-150,-47){\makebox(0,0){\small $\ge_6, \ge_7, \ge_8$}}
\put(-70,55){\line(1,-1){40}}
\put(-30,35){\line(-1,-1){40}}
\put(-50,45){\makebox(0,0){-3}}
\put(-50,5){\makebox(0,0){-2}}
\put(-50,-40){\makebox(0,0){\small $\gg_2 \oplus \gsu(2)$}}
\put(30,70){\line(1,-1){40}}
\put(30,20){\line(1,-1){40}}
\put(70,45){\line(-1,-1){40}}
\put(45,65){\makebox(0,0){-3}}
\put(44,31){\makebox(0,0){-2}}
\put(60, 0){\makebox(0,0){-2}}
\put(50,-40){\makebox(0,0){\small $\gg_2 \oplus \gsu(2)$}}
\put(130,70){\line(1,-1){40}}
\put(130,20){\line(1,-1){40}}
\put(170,45){\line(-1,-1){40}}
\put(145,65){\makebox(0,0){-2}}
\put(144,31){\makebox(0,0){-3}}
\put(160, 0){\makebox(0,0){-2}}
\put(150,-40){\makebox(0,0){\small $\gsu(2) \oplus \gso(7) \oplus \gsu(2)$}}
\end{picture}
\end{center}
\caption[x]{\footnotesize   All possible clusters of intersecting
  curves with self-intersection of each curve $-2$ or below.  For each
cluster the corresponding gauge algebra is noted and the gauge algebra and
matter content are listed in Table~\ref{f:primitives}}
\label{f:clusters}
\end{figure}
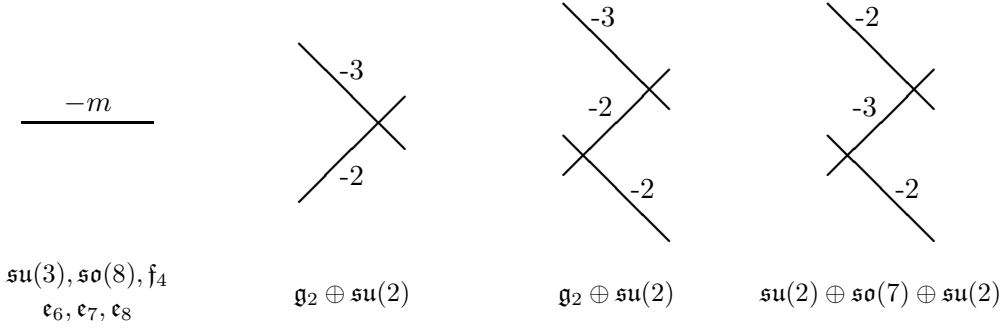

Now we consider clusters consisting of more than two intersecting
irreducible effective divisors.  We consider only clusters where all
divisors have self-intersection $-2$ or less; connection of clusters
by ($-1$)-curves is considered in Section \ref{sec:connecting}.  Since
from the results of the previous section the only pairs that can
intersect are $(-2, -3)$ and $(-2, -2)$, the range of possible
multiple-intersection configurations is limited.  We do not consider
configurations where all curves are $-2$, since these need not carry any
gauge group as discussed above.  So we consider only configurations
that contain at least one ($-3$)-curve.  Simple enumeration of the
possibilities and analysis using the same approach as in the case of
simple pairs shows that there are only two new valid intersecting
clusters.  These are found by adding to intersecting ($-3$)- and
($-2$)-curves another ($-2$)-curve that either intersects the
($-3$)-curve or the ($-2$)-curve.  In the first case, the ($-2$)-curve
carries another $\gsu(2)$ summand; the gauge algebra on the ($-3$)-curve
becomes $\gso(7)$ and there are two sets of half-hypermultiplets,
transforming in the spinor {\bf 8} of $\gso(7)$, with one transforming
in the ${\bf 2}$ of each $\gsu(2)$ and trivially under the other
$\gsu(2)$.  In the second case, the additional curve carries no gauge
group but the degrees of vanishing on the first ($-2$)-curve increase;
monodromy brings the gauge algebra to $\gsp(1) = \gsu(2)$, so we again have
gauge algebra $\gg_2 \oplus \gsu(2)$ with the same matter as in the original
configuration.  These two irreducible geometric units carrying gauge
groups and non-Higgsable matter are depicted in
Figure~\ref{f:clusters} and included in Table~\ref{f:primitives}.  The
detailed derivation of the spectrum on each cluster is given in
subsection \ref{sec:spectra} below.  A systematic analysis shows that
no other irreducible combination of intersecting curves each with
negative self-intersection $-2$ or below
can appear in an F-theory compactification.
The various other possibilities that combine valid pairwise
intersections, including the linear strings of intersecting curves
$(-3, -2, -2, -2), (-2, -3, -2, -2)$ and $(-3, -2, -3)$, combinations
where a single $(-2)$- or ($-3$)-curve is intersected by three $(-2)$- or
($-3$)-curves, as well as a set of three curves with
self-intersections $-3, -2, -2$ that each intersect pairwise, and any
clusters that include these configurations as subclusters, can all be
shown to give elliptic fibrations that are too singular to describe an F-theory compactification
without blowing up points on the base.

For each irreducible cluster of curves appearing in
Table~\ref{f:primitives}, the minimal (non-Higgsable) gauge group and
matter spectrum can be determined either through geometry, as we do in
Section \ref{sec:spectra} below, or by using anomalies.  Each of these
non-Higgsable
gauge + matter configurations has been identified in explicit
string theory constructions, including those with gauge groups
$\gf_4, \gg_2 \oplus \gsu(2)$
and
$\gsu(2) \oplus \gso(7) \oplus \gsu(2)$ \cite{Intriligator}, where the matter content was
determined using anomaly cancellation.
  The analysis here shows that no other
non-Higgsable configurations of gauge groups and matter content can be
required by F-theory geometry.

This completes the analysis of all connected configurations of curves
of self-intersection $-2$ or less that must contain a nonabelian gauge
group or matter content when the model is maximally Higgsed.  We refer
to any of the irreducible clusters and their algebra + matter content (including single
gauge algebra summands without matter such as the $\gf_4$ on a rational
curve of self-intersection $-5$) in Table~\ref{f:primitives} as
``non-Higgsable clusters'', or simply NHC's.

\begin{table}
\begin{center}
\begin{tabular}{| c | c  | c| c | %c | 
c |
}
\hline
Diagram & Algebra & matter & $(f, g, \Delta)$ %& $H_{\rm ch}-V$ 
& $\Delta T_{\rm max}$
\\
\hline
$-3$ & $\gsu(3)$ & 0& $(2, 2, 4)$ %& -8 
&   1/3
\\ 
\hline
$-4$ &$\gso(8)$&0 & $(2, 3, 6)$ %& -28 
& 1
\\ \hline
$-5$ &$\gf_4$ &0 & $(3, 4, 8)$ %& -52 
& 16/9
\\ \hline
$-6$ &$\ge_6$ & 0& $(3, 4, 8)$ %& -78 
& 8/3
\\ \hline
$-7$ &$\ge_7$ &  $\frac{1}{2}${\bf  56}& $(3, 5, 9)$ %& -105 
& 57/16
\\ \hline
$-8$ &$\ge_7$ & 0& $(3, 5, 9)$ %& -133 
&9/2
\\ \hline
$-12$ &$\ge_8$ & 0& $(4, 5, 10)$ %& -248 
&25/3
\\ \hline
$-3, -2$ &  $\gg_2 \oplus \gsu(2)$ & $({\bf 7} + {\bf 1},
\frac{1}{2} {\bf 2})$& $(2, 3, 6), (1, 2, 3)$ %& -9 &  
& 3/8
\\ \hline 
$-3, -2, -2$ &  $\gg_2 \oplus \gsu(2)$ & $({\bf 7} +{\bf 1},
\frac{1}{2} {\bf 2})$& $(2, 3, 6), (2, 2, 4), $ %& - 9 &  
&5/12
\\ 
& & &  (1, 1, 2 ) %& &
&
\\ \hline 
$-2, -3, -2$ &  $\gsu(2) \oplus \gso(7) \oplus \gsu(2)$ & $( {\bf 1},{\bf 8}, \frac{1}{2} {\bf 2})
$ & $(1, 2, 3), (2, 4, 6),$%& -11 &  
&1/2
\\ 
& &$+( \frac{1}{2} {\bf 2},{\bf 8}, {\bf 1})$&  (1, 2, 3) %& &
& 
\\
\hline
\end{tabular}
\end{center}
\caption[x]{\footnotesize  Irreducible geometric components
  (non-Higgsable clusters, or ``NHC's'') consisting
of one or more intersecting curves associated with irreducible
effective divisors each with negative self-intersection.  Each cluster
gives rise to a minimal gauge algebra and matter configuration.}
\label{f:primitives}
\end{table}

\subsection{Higgsing}
\label{sec:Higgsing}

In the analysis above, we identified all combinations of curves that
are forced by their intersection with multiples of $-K$ to carry
certain gauge algebras, in certain cases with given minimal matter
content.  Since the geometry forces these gauge + matter
configurations to appear in the theory, there is no way in which the
matter fields in the last three NHC's in Table~\ref{f:primitives} can
be removed by moving in the moduli space of the theory (without
changing the number of tensor multiplets and moving to a different
base through an extremal transition).  In particular, this means that
the matter in these configurations cannot be Higgsed.  Indeed,
analysis of the matter fields that can be Higgsed for different gauge
algebras shows that Higgsing is impossible in these three situations.
To use fundamentals to Higgs an $\gsu(N)$ gauge theory, two fundamental
matter fields must be simultaneously given expectation values to
implement the Higgsing.  A fundamental + antifundamental is needed to
combine with the broken generators of the $\gsu(N)$ in reduction to
$\gsu(N -1)$ to give the appropriate massive gauge fields.  This can
also be seen from the need to give a second fundamental a VEV to
cancel the D-term constraints in the equations of motion.  Similarly,
for a gauge algebra $\gsu(N) \oplus \G$, one bifundamental field cannot be
Higgsed, two are necessary.  In each of the 3 NHC's described above,
there is a single (half-hyper) transforming under $ \gsu(2) \oplus \G$ as
a fundamental of $\gsu(2)$ and an irreducible representation of $\G$.
Thus, it is clear that none of these configurations can be Higgsed.
On the other hand, $\gso(7)$ can be Higgsed to $\gg_2$ by Higgsing a
single spinor {\bf 8},  and $\gg_2$ can be Higgsed to $\gsu(3)$ by a
single ${\bf 7}$, so these configurations are not non-Higgsable without
the $\gsu(2)$ summands.

We have identified all non-Higgsable configurations that are forced by
the geometry of F-theory from the intersection of the canonical class
with the configurations.  Though it seems plausible physically that
all non-Higgsable configurations are forced in this way from geometry,
this has not been rigorously proven.  One could imagine some kind of
conspiracy in which certain geometries force additional non-Higgsable
matter configurations, for example on curves with positive
self-intersection.  We do not have any argument at this time that
would rule out this possibility in global models, for example, if
the degrees of freedom are highly constrained so that there are no
charged matter field combinations available to break to the minimal
gauge content indicated by the NHC configuration.

One way to show that the maximally Higgsed content of all F-theory
models is captured by the NHC configuration determined from the
divisor structure is to show that all other matter configurations that
may arise in F-theory can be removed by Higgsing.  In most cases this
is fairly straightforward.  For $\gsu(N)$, a single tensor or two-index
antisymmetric representation can be Higgsed, and a variety of Higgsing
mechanisms are known for other common representations and gauge groups
appearing in F-theory \cite{Higgsing-ds}.  As discussed in \cite{0,
Morrison-Taylor}, however, various exotic representations may also be
possible in F-theory.  For example, $3$-index antisymmetric
representations are possible for certain $\gsu(N)$ gauge algebras.  We
have not carried out a completely systematic analysis of all matter
representations --- indeed, it is not yet clear what matter
representations can and cannot be realized in F-theory through
codimension two singularities --- but physically it seems likely that
any Weierstrass model realizing more exotic matter will have free
parameters that remove the matter by Higgsing so that exotic matter
will not appear in generic F-theory constructions over any base.

\subsection{Connecting clusters with ($-1$)-curves}
\label{sec:connecting}

In general, an F-theory base will contain a set of the non-Higgsable
clusters described in Table~\ref{f:primitives}.  Some or all of these
clusters will be connected by ($-1$)-curves intersecting curves from the
clusters.  Note that not all clusters can be connected by ($-1$)-curves.
For example, a $(-4)$-curve $A$ cannot be connected to a ($-5$)-curve $B$
by a ($-1$)-curve $C$, since the degrees of vanishing of $f, g, \Delta$
on $B$ and $C$ would have to be 3, 4, 8, and 2, 2, 4 respectively, so
that the intersection point between these curves would need to be
blown up for a good F-theory base.  A similar analysis can be used to
identify all ways in which ($-1$)-curves connect NHC's.

One simple rule for when a ($-1$)-curve can connect a set of NHC's is
that any set of curves 
can be connected by a ($-1$)-curve $A$
when the total (including
intersection multiplicities) of the degrees of $f, g, \Delta$ on the
curves intersecting $A$ is less than or equal to 4, 6, 12
respectively.  To prove this consider a configuration where $A$
intersects curves $C_i$ with multiplicity $p_i$.
Any multiple of $-K$ can then be written as
\begin{equation}
-nK = aA + \sum_{i}c_iC_i + X
% \label{eq:}
\end{equation}
where $X \cdot A \geq 0, X \cdot C_i \geq 0$.  The intersection with
$A$ is
\begin{equation}
-nK \cdot A = n = -a + \sum_{i}c_i p_i + X \cdot A \,.
% \label{eq:}
\end{equation}
The positivity of $X \cdot A$ then implies
\begin{equation}
a \geq \sum_{i}c_ip_i-n \,.
\label{eq:a-condition}
\end{equation}
When 
\begin{equation}
\sum_{i} c_i p_i\leq n
\label{eq:cp}
\end{equation}
then this condition can be satisfied with $a = 0$.  Thus, any
configuration of curves satisfying the condition above for $f, g,
\Delta$ ({\it i.e.}, satisfying \eq{eq:cp} for the decomposition of
$-nK$ for $n = 4, 6, 12$) can be connected by a ($-1$)-curve $A$ that is
not contained in any multiple of $-K$ and does not affect the gauge
group or matter content of the theory.

As an example of this condition, a ($-1$)-curve can connect two 
($-4$)-curves and does not carry any degree of vanishing for $f, g, \Delta$.
On the other hand, a ($-1$)-curve cannot connect a $(-4)$-curve to a 
($-5$)-curve, as mentioned above.  A ($-1$)-curve can only connect a $-12$
curve to another ($-1$)-curve or a ($-2$)-curve that does not carry a
gauge group.

There are a limited number of ways in which a ($-1$)-curve can intersect a
single NHC.  Using the analysis of which pairs of curves can intersect
from Section \ref{sec:pairs}, we know that a ($-1$)-curve can intersect
a curve of self-intersection $-5$ or less only once, and a curve of
self-intersection $-4, -3,$ or $-2$ either once or twice.  Performing a
case-by-case analysis shows that there are 31 distinct ways in which a
($-1$)-curve can intersect a single NHC in a consistent F-theory base.
It is then possible to systematically analyze all possible
combinations of NHC's that can be intersected by a single ($-1$)-curve.
The details of this analysis are presented in the Appendix.

While the relation \eq{eq:cp} holding for $n = 4, 6,$ and 12
is a sufficient condition for an intersection configuration to be
allowed, the opposite is not true.  In almost all cases, when
\eq{eq:cp} is violated for $n = 12$ the configuration of NHC's
connected by a ($-1$)-curve becomes singular and is not allowed.  The
only exceptions to this are when the ($-1$)-curve $A$ intersects a 
($-5$)-curve and the ($-3$)-curve of either a $(-3, -2)$ NHC or a $(-3, -2,
-2)$ NHC.  In these cases, $\sum_{i}c_i = 14$ for $n = 12$ and the vanishing degree
of $\Delta$ on $A$ is 2, but there is no 4, 6, 12 singularity at any
point.  There is, however, no nonabelian gauge group factor carried
on $A$, and the algebra structure of the connected NHC's is unchanged.
Thus, all configurations of NHC's connected by ($-1$)-curves give rise
to the same gauge algebra and matter content as if they were isolated.

There are also a number of marginal cases, where a ($-1$)-curve $A$
can connect NHC's in such a way that $A$ must appear in $f$ or $g$
but need not contribute to the discriminant $\Delta = -12K$ (and
therefore again need not carry any gauge group).  For example, consider the
case where $A$ connects a ($-3$)-curve to a $(-5)$-curve.  For $g,
\Delta$, \eq{eq:cp} is satisfied, as can be readily confirmed from the
data in Table~\ref{f:primitives}.  On the other hand, $f$ has
vanishing degrees of 2 and 3 on the $(-3)$- and $(-5)$-curves respectively
and thus $a \geq 1$ from \eq{eq:a-condition}.  

It is also not the case that any NHC combination can be connected
where \eq{eq:cp} is satisfied for $n = 12$.  As a counterexample,
consider a ($-1$)-curve connecting four $(-3, -2)$ type NHC's by
intersecting each along the corresponding ($-2$)-curve $C_1, \ldots
C_4$.  In this case, \eq{eq:cp} is satisfied for $f$ and $\Delta$ but
not for $g$.  In fact, this is not a valid configuration for an
F-theory base.  To see this, write
\begin{equation}
-6K = aA + \sum_{i}c_iC_i + \sum_{i} d_iD_i  +X\,
% \label{eq:}
\end{equation}
where $D_i$ are the four ($-3$)-curves.
The intersection product with $A$ gives
\begin{equation}
-6K \cdot A = 6 = \sum_{i} c_i-a + X \cdot A \,,
% \label{eq:}
\end{equation}
so $a \geq \sum_{i} c_i-6$.
The intersection product with $C_i$ gives
\begin{equation}
-6K \cdot C_i = 0 = d_i + a-2c_i + X  \cdot C_i \,,
% \label{eq:}
\end{equation}
so (since $d_i \geq 3$), $c_i \geq (3 + a)/2$.  Combining these two
equations gives $a \geq 2a$, and since $c_i \geq 2$  it must also be
the case that $a > 0$.  These equations cannot be simultaneously
satisfied so there is no good F-theory base with this geometry.

The complete analysis of all ways in which a single ($-1$)-curve can
connect to one or more NHC's is summarized in the Appendix.  The total
number of distinct ways that a ($-1$)-curve can connect $k$ NHC's is:
\vspace*{0.1in}

\noindent
31 configurations with a ($-1$)-curve intersecting 1 NHC

\noindent
100 configurations with a ($-1$)-curve intersecting 2 NHC's

\noindent
46 configurations with a ($-1$)-curve intersecting 3 NHC's

\noindent
6 configurations with a ($-1$)-curve intersecting 4 NHC's
\vspace*{0.1in}

There are no ways of connecting more than 4 NHC's with a single
$(-1)$-curve.  This gives a total of 183 ways in which a ($-1$)-curve
can connect to a combination of NHC's.  In all of these
configurations, the ($-1$)-curve carries no nonabelian gauge group,
and the total gauge group is the product of factors from the NHC's.

\subsection{Minimal spectra on irreducible clusters}
\label{sec:spectra}

We have determined above the neccessary orders of vanishing of $f$, $g$, 
$\Delta$ along the components of each irreducible cluster, but we have
not yet explained how to determine the gauge algebra and charged matter content
in each case.  We  do that in this section, using the methods of
\cite{Grassi-Morrison-2}.

The data of the orders of vanishing of $f$, $g$ and $\Delta$ along a curve $C$
must in many cases be
supplemented by some additional information in order to
determine the gauge algebra.  In \cite{Grassi-Morrison-2}, this is
expressed in terms of a ``monodromy cover''
 of $C$ given by a polynomial equation
$\mu(\psi)=0$ of degree $2$ or $3$, and it must be determined whether
this is an irreducible cover, and if reducible, how many components
it has.  When the monodromy cover has degree $2$ there is a natural
condition which determines this: one needs to check if the ramification
divisor (which is the divisor of zeros of the discriminant  of $\mu(\psi)$)
 is divisible by two or not.  When the monodromy cover has degree $3$,
however, the situation is much more subtle; we will return to this case
shortly.

Closely related to the monodromy 
analysis is one of the two types of contributions
to the charged matter content of the F-theory compactification: 
the ``non-local matter.''
The non-local matter consists of ${\rm genus}(C)$ adjoint hypermultiplets, together
with ${\rm genus}(\widetilde{C})$ copies of another representation if there is a component $\widetilde{C}$
of the monodromy cover which is not isomorphic to $C$.  In addition to
the non-local matter, there are ``local'' contributions to the matter
representation from each zero of the residual discrimimant $\Delta_0 = 
(\Delta/z^n)|_C$ (although a local contribution may be trivial if the 
corresponding zero is also in the ramification divisor of the monodromy cover).

In the case of an isolated curve $C$ which is an NHC, we write 
$-12K = mC + X$ where $m$ is the order of vanishing of $\Delta$, so that
$\Delta_0 = X\cdot C$.  It is easy to check that $X\cdot C=0$ except in 
two cases.
For a
$(-7)$-curve $C$, we write $-12K = 9C + X$ and find that $X\cdot C =3$.
In this case, $\Delta_0 = ((f/z^3)|_C)^3$, so $\Delta_0$ has a single zero
of multiplicity three: it corresponds to a $\frac12\mathbf{56}$ 
hypermultiplet of $\mathfrak{e}_7$, according to \cite{Grassi-Morrison-2}.

Similarly, for a $(-5)$-curve $C$,
we write $-12K = 8C + X$ and find that $X\cdot C=4$.
In this case, according to \cite{Grassi-Morrison-2}, we have
$\Delta_0 = (\operatorname{disc}(\mu(\psi)))^2$,
so the ramification divisor of the monodromy cover has two zeros.
If those zeros are not distinct, then the cover splits into two
components, the gauge algebra
is $\mathfrak{e_6}$, and the matter representation is a single $\mathbf{27}$
hypermultiplet.  From the geometry side, we can vary coefficients in the
equation to guarantee that the zeros of $\operatorname{disc}(\mu(\psi))$
become distinct; from the physics side,
we can Higgs the $\mathbf{27}$. Thus, from either approach we see that
this is not the correct description of the NHC.

The alternative is that the two zeros of the ramification divisor are distinct.
Then the gauge algebra is $\mathfrak{f}_4$, 
we have ${\rm genus}(\widetilde{C})=0$ so there is no non-local matter, and the zeros of
$\operatorname{disc}(\mu(\psi))$ do not contribute to the local matter either.
Thus, we find an NHC as stated in Table~\ref{f:primitives}: the gauge algebra
is $\mathfrak{f}_4$ and there is no charged matter.

In all other cases of an NHC with a
single curve, $X\cdot C=0$ and there is no possibility of localized matter.
For $C^2=-3$, $-4$, or $-6$, we must determine the monodromy cover in order to
determine the gauge algebra.  However, in all of these cases, $X\cdot C=0$
implies that the monodromy polynomial $\mu(\psi)$ has constant coefficients
and so it must factor completely.  This leads to the maximum possible
gauge algebra in these three cases, namely $\mathfrak{su}(3)$,
$\mathfrak{so(8)}$, and $\mathfrak{e}_6$.  Moreover, there cannot be
any non-local matter in these cases because ${\rm genus}(C)=0$.

We note in passing that this discussion illuminates our earlier
explanation of why curves $C$ with $C^2= -9$, $-10$, or $-11$
cannot be NHC's.  In these cases, we write $-12K=10C+X$ and find
that $X\cdot C = 6$, $4$, or $2$.  But any nonzero value of $X\cdot C$
with gauge algebra $\mathfrak{e}_8$ 
leads to a point of intersection where the multiplicities
exceed $(4,6,12)$ and so such curves are not allowed on an F-theory base.

We return now to a closer analysis of monodromy covers of degree $3$.
The only case in which a monodromy cover of degree $3$ occurs is 
when $f$ vanishes to order at least 2, $g$ vanishes to order at least $3$,
and $\Delta$ vanishes to order exactly $6$.  In this case, the covering polynomial is
\begin{equation}
 \mu(\psi) = \psi^3 + (f/z^2)\psi + (g/z^3),
\end{equation}
where $z=0$ is a local defining equation for $C$.  The discriminant of
this polynomial is 
\begin{equation}
\operatorname{disc}(\mu) = 4(f^3/z^6) + 27 (g^2/z^6) = \Delta/z^6=\Delta_0.
\end{equation}
Let us observe some
things about the three possible cases, in which the monodromy cover has various
numbers of components.

If the monodromy cover has three components, then the covering polynomial
factors as
\begin{equation}
\mu(\psi) = (\psi-a)(\psi-b)(\psi-c),
\end{equation}
where $a+b+c=0$.  This is the case of gauge algebra $\mathfrak{so}(8)$.
The discriminant of this polynomial is
\begin{equation}
\Delta_0 = (a-b)^2(a-c)^2(b-c)^2,
\end{equation}
which is a perfect square.  Moreover, the three factors of $\sqrt{\Delta_0}$
are associated to three different local contributions to the charged matter
representation: each zero of $a-b$ is associated to an $8_v$, each
zero of $a-c$ is associated to an $8_+$ and each zero of $b-c$ is 
associated to an $8_-$ (up to the permutation among these representations
induced by triality of $\mathfrak{so}(8)$).

If the monodromy cover has two components, then the polynomial $\mu(\psi)$ factors as
\begin{equation}
 \mu(\psi) = (\psi - a) (\psi^2 + d\psi + e)
\end{equation}
(where $a-d=0$),
and the quadratic factor must define an irreducible cover, i.e.,
\begin{equation}
 \operatorname{disc}(\psi^2 + a\psi + b) = a^2-4b
\end{equation}
must not be a square.  This is the case of gauge algebra
$\mathfrak{so}(7)$.  In this case, we can write
\begin{equation}
 \Delta_0 = - (a^2-4b)(2a^2-b)^2,
\end{equation}
that is, $\Delta_0 = \alpha \beta^2$ with $\alpha$ not a square.

The zeros of $\beta$ are associated to spinor representations $\mathbf{8}$
of $\mathfrak{so}(7)$.  The zeros of $\alpha$ do not directly determine
matter, but they do determine the genus of the nontrivial component 
$\widetilde{C}$ in the monodromy cover, and there are ${\rm genus}(\widetilde{C})$
copies of the vector representation $\mathbf{7}_v$ of
$\mathfrak{so}(7)$.

Finally, if the polynomial $\mu(\psi)$ is irreducible, the gauge algebra
is $\mathfrak{g}_2$ and there are no
local contributions to the matter.  However, the zeros of the discriminant
will determine the genus of the cover $\widetilde{C}$, and there are
${\rm genus}(\widetilde{C})$ copies of the $7$-dimensional representation.

Let us now use these facts to finish our analysis of the NHC's in the last
three lines of Table~\ref{f:primitives}.  In all three cases, we have
a curve $C_1$ of self-intersection $-3$ with $(f,g,\Delta)$ multiplicities
of $(2,3,6)$ or $(2,4,6)$.  
Thus, we are in the situation with a monodromy cover of
degree $3$.  Moreover, if we write $-12K = 6C_1 + X$, then $X\cdot C_1 = 6$
so there is room for some charged matter.  
If the factor of the gauge algebra corresponding to $C_1$ 
is $\mathfrak{so}(8)$,
then $X\cdot C_1$ will be twice a divisor of degree $3$, and the
matter representation will be $\mathbf{8}_v \oplus \mathbf{8}_+ \oplus 
\mathbf{8}_- $.  (Note that the matter may be charged with respect to
other summands of the gauge algebra as well: we will come back to that point.)
If the gauge algebra is $\mathfrak{so}(7)$,
then $X\cdot C_1$ takes the form $D_1+2D_2$ with each $D_i$ having degree $2$,
and the matter consists of two copies of $\mathbf{8}$ (because in
this case, ${\rm genus}(\widetilde{C})=0$).  And finally, if the gauge algebra
is $\mathfrak{g}_2$ then the charged matter represention is 
${\rm genus}(\widetilde{C})=1$
copy of the $7$-dimensional representation.

The NHC clusters $(-3,-2)$ and $(-2,-3,-2)$  have a second curve $C_2$
of self-intersection $-2$ with $(f,g,\Delta)=(1,2,3)$
which meets $C_1$.  We can write
\begin{equation}
-12K = 6C_1 + 3C_2 + Y
\end{equation}
and we see the previous $X\cdot C_1$ break up into $3 P + Y\cdot C_1$
where $P$ is the intersection point of $C_1$ and $C_2$.  Thus, if
the gauge algebra factor for $C_1$ is $\mathfrak{so}(7)$ we must have
$Y\cdot C_1 = P + Q_1 + Q_2$ and if it is $\mathfrak{so}(8)$ we must
have $Y\cdot C_1 = P + 2Q$.  (Both of these are non-generic, so we expect
geometrically to get to $\mathfrak{g}_2$ by choosing $f$ and $g$
generically.)  

On the other hand, $(6C_1 + Y)\cdot C_2 = 6P + Y\cdot C_2$ should be
three times a divisor (according to \cite{Grassi-Morrison-2} for type III)
and as is easy to calculate, $Y\cdot C_2=0$, so the localized matter
is associated with the divisor $D=2P$.  The spectrum consists of $2\deg{D}$
fundamentals, i.e.\ $4$ fundamentals,
or $8$ half-fundamental hypermultiplets.  Thus, the combined matter
associated to the point $P$ adds up to
\begin{equation}
(8,\frac12\mathbf{2})
\end{equation}
as a representation of $\mathfrak{g}(C_1)\oplus \mathfrak{su}(2)$,
where $8$ is an appropriate $8$-dimensional representation of
$\mathfrak{so}(8)$, $\mathfrak{so}(7)$ or $\mathfrak{g}_2$.  (In the
latter case, the representation can be written $\mathbf{7}\oplus\mathbf{1}$.)

In the $(-3,-2)$ case, if the gauge algebra is larger there will be
additional matter of the form $(8,\mathbf{1})$ (one such multiplet for
$\mathfrak{so}(7)$ and two from $\mathfrak{so}(8)$) but these can be
Higgsed, leaving the NHC with gauge algebra $\mathfrak{g}_2\oplus\mathfrak{su}(2)$ and matter $(\mathbf{7}\oplus\mathbf{1},\frac12\mathbf{2})$.

In the $(-2,-3,-2)$ case, there is an additional curve $C_0$ leading to
a second local matter contribution of the form $(8,\frac 12(\mathbf{2}))$,
this time charged under the $\mathfrak{su}(2)$ associated to $C_0$.
Thus, in this case the minimum gauge algebra which can occur is 
$\mathfrak{so}(7)$ and if $\mathfrak{so}(8)$ occurred there would be
an extra multiplet neutral under both $\mathfrak{su}(2)$'s, which could be
Higgsed.  The conclusion is that the minimum
gauge algebra is $\mathfrak{su}(2)\oplus
\mathfrak{so}(7)\oplus \mathfrak{su}(2)$.

Finally, in the $(-3,-2,-2)$ case, the curve $C_2$ adjacent to $C_1$
is a $(-2)$-curve with $(f,g,\Delta)=(2,2,4)$, and the third curve $C_3$
with $(f,g,\Delta)=(1,1,2)$
makes no contribution to the gauge algebra.  In this case, we write
\begin{equation}
-12K = 6C_1 + 4C_2 + 2C_3 + Z,
\end{equation}
and we find $Z\cdot C_1=2$ while $Z\cdot C_2=Z\cdot C_3=0$.
According to \cite{Grassi-Morrison-2}, 
the residual discriminant for the $C_2$ component satisfies
$\Delta_0=(\operatorname{disc}(\mu(\psi))^2$ in this type IV case,
so it has divisor $3P+Q$ where $P=C_1\cap C_2$ and $Q=C_2\cap C_3$.
In particular, the ramification divisor is not even, so the gauge
algebra is $\mathfrak{su}(2)$ rather than $\mathfrak{su}(3)$.
%As in \cite{Grassi-Morrison-2},
Anomaly cancellation implies that the charged matter for this
$\mathfrak{su}(2)$ is $4$  fundamentals, so we have a very similar spectrum
to the $(-3,-2)$ case.  Note that in this case as well, varying coefficients
in $f$ and $g$ or Higgsing the extra $8$-dimensional representations
(which are neutral under $\mathfrak{su}(2)$) should reduce that gauge
algebra summand to $\mathfrak{g}_2$.

\section{Bounding the number of tensors}
\label{sec:bounds}

From the analysis of the previous section, we expect that all F-theory
models will have a maximally Higgsed gauge group and matter content
that decomposes into factors associated with the non-Higgsable clusters
(NHC's) listed in Table~\ref{f:primitives}.  This gives a simple way
of classifying theories according to gauge algebra and matter content.
In this section we show that a theory with any given
combination of NHC's has a
maximum number of tensors.

\subsection{Theories without vector multiplets}
\label{sec:no-vectors}
\label{sec:bound-simple}

We begin by considering F-theory constructions of theories without
gauge groups.  Such models arise as generic (maximally Higgsed)
theories on F-theory bases with no irreducible effective divisors
that are rational curves with $C \cdot C < -2$. 
Thus, on such a base $-K \cdot C \geq 0$ for all rational curves.
The analysis of more general F-theory models follows a similar logic,
though the details are more complicated; we consider general models in
the following section.

We assert that for any F-theory model without vector multiplets the
number of  tensor multiplets is bounded above by 
\begin{equation}
T < 10   \,, \;\;\;\;\;\makebox{if the gauge group $G$ is trivial.}
\label{eq:simple-bound}
\end{equation}

This can be proven from F-theory
as follows: decompose $-K$ into a sum of irreducible
effective curves as
\begin{equation}
-K = \sum_{i} l_i C_i
\label{eq:k-summation}
\end{equation}
If $K \cdot K < 0$ then for some $i$, $-K \cdot C_i < 0$ and there is a
nonabelian gauge group factor.  Thus, for theories with trivial gauge
group $K \cdot K = 9-T\geq 0$ so $T < 10$.

Another way to understand this bound is from the
gravitational anomaly cancellation condition
\begin{equation}
H-V= 273-29T
% % \label{eq:}
\end{equation}
In this relation, $H$ is the total number of hypermultiplets, $V$ is
the number of vector multiplets, and $T$ is the number of tensors.
For theories with no vector multiplets, we have $H = 273-29T$, which
must be positive, so $T \leq 9$.  The argument based on
(\ref{eq:k-summation}) carries the same information as the
gravitational anomaly bound, though from a rather different geometric
perspective.  Note that the first argument depends upon an F-theory
realization, while the second argument only depends upon quantum
consistency of the supergravity theory.  On the other hand, the second
argument assumes that there are no $U(1)$ factors in the gauge group,
while the first argument is independent of the existence of $U(1)$ factors.
$U(1)$ factors add significant subtleties to analysis from the
supergravity point of view, as discussed in \cite{Park-Taylor, Park}.

There is a complete classification of the possible F-theory bases
without nonabelian gauge group factors in the generic maximally
Higgsed phase.  Surfaces on which all rational curves are ($-1$)-curves
are del Pezzo surfaces.  The del Pezzo surface ${\rm dP}_k$ is
constructed by blowing up $k$ generic points on $\P^2$.  Surfaces
containing rational curves with self intersection $-1$ and $-2$ are
{\it generalized del Pezzo} surfaces \cite{777}.  Generalized del Pezzo surfaces
are limits in the moduli space of ordinary del Pezzo
surfaces.  Generalized del
Pezzo surfaces can be classified by the intersection structure of the
($-2$)-curves.  For a generalized del Pezzo surface with $h^{1, 1} (B) = T
+ 1 = k + 1,$ this intersection structure must form a sub-root lattice
of the affine root lattice $\hat{E}_k$ \cite{pinkham}.
Generalized del Pezzo surfaces all have $T \leq 9$, in agreement with
the bound \eq{eq:simple-bound}.

\subsection{Bounding the number of tensors for a given gauge algebra} 
\label{sec:bound-general}

We can now generalize the argument from Section
\ref{sec:bound-simple} to an
arbitrary F-theory base.  Any F-theory base has a system of effective
irreducible divisors characterized by a set of disjoint irreducible
components (NHC's)
from Table~\ref{f:primitives}.  These components are
connected by curves of self-intersection $-1$  that carry no
gauge group (with additional possible clusters of ($-2$)-curves
appearing that carry no gauge group as in the generalized del Pezzo
models discussed above).  For each NHC
component, the bound on the number of tensors increases.  For example,
consider an $\gso(8)$ summand associated with an isolated $(-4)$-curve $C$.
Since $-12K$ must include $6$ copies of $C$, and satisfies $-12K  \cdot
C = -24$,
we have
\begin{equation}
-12K = 6C + X
\;\rightarrow \; 144 K^2 = 36C^2 + X^2 = -144 + X^2 \,,
% % \label{eq:}
\end{equation}
where we have used $X \cdot C = 0$.
Following the same reasoning as above, if the only
gauge algebra summand is the single $\gso(8)$, then $X^2 \geq 0$, so
$K^2 > -1$.  We have proven that
\begin{equation}
\G = \gso(8) \; \Rightarrow \; T \leq 10 \,.
% % \label{eq:}
\end{equation}
Each additional gauge algebra summand of $\gso(8)$ in the maximally Higgsed
theory on a given base thus raises the bound on $T$ by 1.

The same general argument can be carried out for each of the other minimal
components.   Consider an isolated ($-m$)-curve $C$ that appears in
$\Delta$ with multiplicity $c = \lceil 12 (m-2)/m \rceil$.  We have $C^2 =
-m, -K \cdot C = 2-m$, so writing $-12K = cC + X$ we have $(-12K)
\cdot C = cC \cdot C + X \cdot C$ and $X\cdot C = 24 -12m+ cm$.  We
can then substitute into
\begin{equation}
144K^2 =  c^2 C^2 + 2c X \cdot C + X^2
=  X^2 -c (24 m-mc-48) \,.
% % \label{eq:}
\end{equation}
Thus, each such component in the maximally Higgsed gauge group increases
the allowed number of tensors by 
\begin{equation}
\Delta T_{\rm max} =\frac{1}{144}c (24 m-mc-48)
 =\frac{1}{144}(24 m-m \lceil 12 (m-2)/m \rceil-48)\lceil 12 (m-2)/m \rceil\,.
% \label{eq:}
\end{equation}
These are the
terms listed in the last column of Table~\ref{f:primitives}.

A similar contribution to the bound on $T$ arises for non-Higgsable
clusters containing more than one curve of self intersection $-2$ or
below.  A simple calculation along the lines of the above shows that
the extra contribution to the bound on $T$
from a general NHC is the sum of the terms from each curve, plus an
additional contribution from each intersection of $2c_ic_jC_i \cdot
C_j$, where we expand $-12K = \sum_{i}c_iC_i + X$.
We have tabulated the increase in the bound on the number of tensors
for each non-Higgsable cluster in Table~\ref{f:primitives}.

\subsection{Bounds on linear chains of curves}

As an example of how the bounds described in the previous section can
be combined with knowledge of possible NHC configurations to limit the
range of possible F-theory bases, we consider
linear chains of effective irreducible divisors with negative
self-intersection.  This gives a simple class of configurations of
curves with large gauge groups where the upper bound on the number of
tensors and NHC's is clearly evident.  We consider in particular
linear chains of divisors $C_i$ each of negative self-intersection,
where the only nonzero intersections between distinct divisors is $C_i
\cdot C_{i + 1} = 1$.  We will find it particularly interesting to
consider linear chains that consist of a repeated pattern of divisors.
Such a pattern, for example, appears in the F-theory base for the
model with the largest known value of $T$
\cite{Aspinwall-Morrison-instantons}.

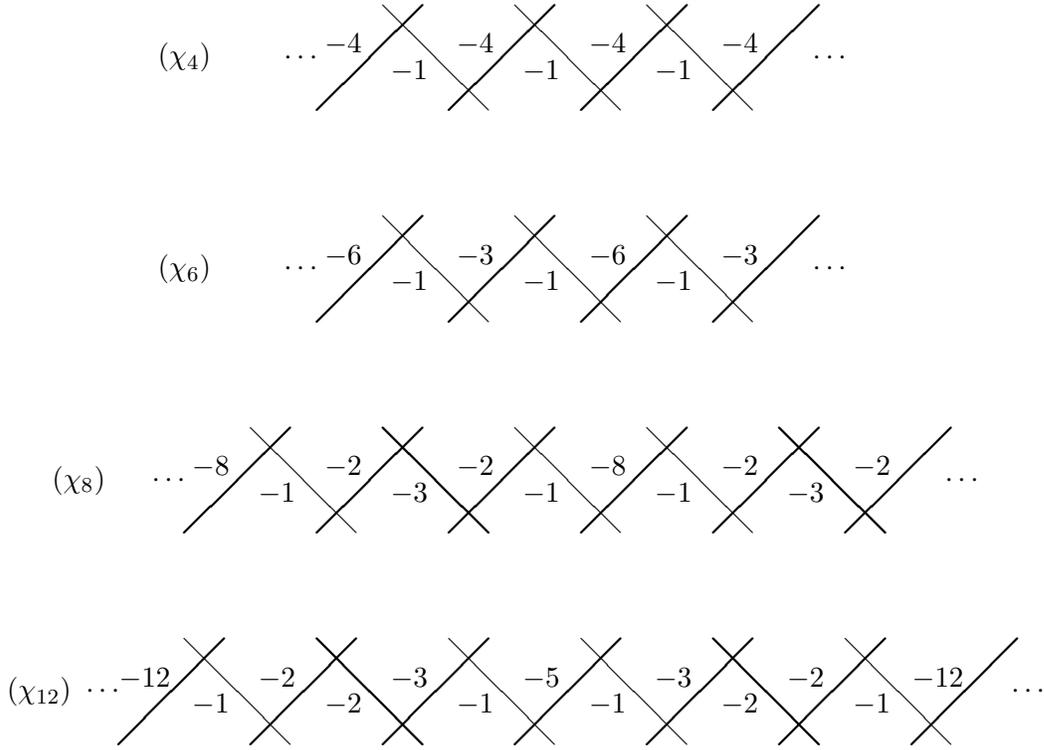
\begin{figure}
\begin{center}
\begin{picture}(200,300)(- 100,- 150)
\multiput(-70,60)(50,0){3}{\line(1,-1){40}}
\multiput(-70,140)(50,0){3}{\line(1,-1){40}}
\multiput(-120,-20)(50,0){5}{\line(1,-1){40}}
\multiput(-145,-100)(50,0){6}{\line(1,-1){40}}
\thicklines
\multiput(-95,100)(50,0){4}{\line(1,1){40}}
\multiput(-95,20)(50,0){4}{\line(1,1){40}}
\multiput(-145,-60)(50,0){6}{\line(1,1){40}}
\multiput(-170,-140)(50,0){7}{\line(1,1){40}}
\multiput(-70,-20)(150,0){2}{\line(1,-1){40}}
\multiput(-95,-100)(150,0){2}{\line(1,-1){40}}
\multiput(-85,125)(50,0){4}{\makebox(0,0){$-4$}}
\multiput(-60,115)(50,0){3}{\makebox(0,0){$-1$}}
\multiput(-85,45)(100,0){2}{\makebox(0,0){$-6$}}
\multiput(-35,45)(100,0){2}{\makebox(0,0){$-3$}}
\multiput(-60,35)(50,0){3}{\makebox(0,0){$-1$}}
\multiput(-135,-35)(150,0){2}{\makebox(0,0){$-8$}}
\multiput(-110,-45)(150,0){2}{\makebox(0,0){$-1$}}
\multiput(-85,-35)(150,0){2}{\makebox(0,0){$-2$}}
\multiput(-60,-45)(150,0){2}{\makebox(0,0){$-3$}}
\multiput(-35,-35)(150,0){2}{\makebox(0,0){$-2$}}
\multiput(-10,-45)(150,0){1}{\makebox(0,0){$-1$}}
\multiput(-160,-115)(300,0){2}{\makebox(0,0){$-12$}}
\multiput(-135,-125)(250,0){2}{\makebox(0,0){$-1$}}
\multiput(-110,-115)(200,0){2}{\makebox(0,0){$-2$}}
\multiput(-85,-125)(150,0){2}{\makebox(0,0){$-2$}}
\multiput(-60,-115)(100,0){2}{\makebox(0,0){$-3$}}
\multiput(-35,-125)(50,0){2}{\makebox(0,0){$-1$}}
\multiput(-10,-115)(300,0){1}{\makebox(0,0){$-5$}}
\put(-145,120){\makebox(0,0){($\chi_4$)}}
\put(-145,40){\makebox(0,0){($\chi_6$)}}
\put(-185,-40){\makebox(0,0){($\chi_{8}$)}}
\put(-200,-120){\makebox(0,0){($\chi_{12}$)}}
\put(-100,120){\makebox(0,0){$\cdots$}}
\put(100,120){\makebox(0,0){$\cdots$}}
\put(-100,40){\makebox(0,0){$\cdots$}}
\put(100,40){\makebox(0,0){$\cdots$}}
\put(-175,-120){\makebox(0,0){$\cdots$}}
\put(175,-120){\makebox(0,0){$\cdots$}}
\put(-150,-40){\makebox(0,0){$\cdots$}}
\put(150,-40){\makebox(0,0){$\cdots$}}
\end{picture}
\end{center}
\caption[x]{\footnotesize Periodic linear chains of divisors  with
  simple gauge algebras.  Chains shown are all those that satisfy a
  local maximality condition  on the self-intersection numbers of the
  divisors.  Bounds on the number of tensors given the gauge group
  place a limit on the size of such chains.}
\label{f:chains}
\end{figure}

Consider for example a periodic chain of divisors with
self-intersections 
\begin{equation}
(\ldots, -4, -1, -4, -1, -4, \ldots)
% \label{eq:}
\end{equation}
(See
Figure~\ref{f:chains} (a)).  For the moment we simply consider an
idealized infinite divisor chain; boundary conditions for finite
chains will be discussed shortly.  This chain of divisors has several
properties:

\begin{enumerate}
\item Each link in the chain is allowed in an F-theory model ({\it
  i.e.}, each link is either part of an NHC or a ($-1$)-curve connecting
  NHC's in a fashion allowed by the rules in
  Table~\ref{t:intersections} in the Appendix)
\item The chain is (locally) {\it maximal}, in the sense that no
  self-intersection number of any link in the chain can be increased
  while maintaining property (1).
\item The chain can be reduced by blowing down successive ($-1$)-curves
  until all nonabelian gauge groups are removed without separating the chain.
\end{enumerate}

We call a chain with the first two properties a {\it maximal linear
divisor chain}.  To verify the third property, note that when a $(-1)$-curve $A$ connecting curves $B, C$ of self-intersection $-m$ and $-n$
is blown down, the ($-1$)-curve is replaced by a single point of
intersection of $B$ and $C$, and the self-intersections of these
curves become $-m +1$ and $-n +1$.  In the specific example of the
chain (\ref{eq:chain-4}), blowing down each of the ($-1$)-curves gives a
new chain of the form $( \ldots, -2, -2, -2, \ldots)$ with no curves
of self-intersection below $-2$ and hence no gauge group.

Now consider the gauge algebra on the finite chain 
\begin{equation}
(-1, -4, -1, \ldots, -1, -4, -1)
\label{eq:chain-4}
\end{equation}
with $N$ $(-4)$-curves.  From Table~\ref{t:Kodaira} this combination of
NHC's gives $N$ nonabelian gauge algebra summands $\gso(8)$.  From the
analysis in the previous section, the  number of tensors in a
model with this gauge algebra is bounded by $T \leq 9 + N$.  Blowing
down all ($N + 1$) of the ($-1$)-curves in the chain gives a chain of
length $N$ containing only ($-2$)-curves.  For this to be an F-theory
model there must be a boundary condition that enables this chain to be
blown down further to get to either $\F_m$ or $\P^2$.  Each blow-down
removes one ($-2$)-curve.  So the number of blow-downs necessary to
reduce the original chain to one with a single curve of negative
self-intersection is of order ${\cal O} (2 N)$.  This means that the
original chain had a number of tensors of $T \sim 2 N$.  Comparing to
the bound we have
\begin{equation}
T \sim 2 N \leq 9 + N \,.
% \label{eq:}
\end{equation}
This means that the chain of $\gso(8)$ NHC's cannot contain more than on
the order of 9 $(-4)$-curve components.  This can also be seen by noting that
the number of $-2$ factors in a linear chain in a model with no
nonabelian gauge groups is $< 9$, from the discussion of generalized
del Pezzo surfaces above.  A more precise bound on $N$ for the chain
(\ref{eq:chain-4}) requires a specific choice of boundary condition
that allows the chain to be blown down to a surface $\F_m$.  By
choosing this boundary condition properly it can be shown that the
value $N= 9$ can be realized.  The basic idea is to terminate the
chain on both ends with the $(-4)$-curves, including two extra  ($-1$)-curves connecting to the next-to-last $(-4)$-curves on each end.  By
blowing down from one end and leaving a single terminal $-4$ intact,
this gives a configuration that can be blown down to $\F_4$ with $16$
blow-downs, giving $T = 17 < 9 + 9 = 18$.  This example and other
related examples are described in more detail in the context of a
complete analysis of toric bases in a paper that will appear as a
sequel to this work \cite{mt-toric}.

Now let us consider other possible periodic maximal linear divisor
chains.  There are only a few other possibilities that satisfy
conditions (1) and (2) above.  These are the periodic chains
\begin{itemize}
\item[$\chi_6$:] $(\ldots, -6, -1, -3, -1, -6, -1, -3, \ldots)$
\item[$\chi_8$:] $(\ldots, -8, -1, -2, -3, -2, -1, -8, -1, -2, -3, -2, \ldots)$
\item[$\chi_{12}$:] $(\ldots, -12, -1, -2,  -2, -3,-1, -5, -1, -3, -2, -2, -1, -12, \ldots)$
\end{itemize}
These chains can be generated by starting with a  NHC containing a
single ($-m$)-curve with $m = 6, 8, 12$ and connecting iteratively to
the curve of the most negative possible self-intersection.  The same
algorithm generates the sequence (\ref{eq:chain-4})
$(\chi_4)$ when begun with $m
= 4$, and sequence $\chi_{12}$ above when begun with $m = 5$.  For each of the
chains  $\chi_n$ that satisfy conditions (1) and (2), condition (3) is
satisfied as well.  In each case, iteratively blowing down all possible ($-1$)-curves leads to a sequence of connected ($-2$)-curves carrying no gauge
group, just as in the case of the sequence (\ref{eq:chain-4}).  In
fact, blowing down the ($-1$)-curves appearing in $\chi_8$ gives the chain
$\chi_6$, and blowing down the ($-1$)-curves in $\chi_6$ gives the chain
$\chi_4$. Blowing down the ($-1$)-curves in $\chi_{12}$ gives the
chain $(-10, -1, -2, -2, -3, -2, -2, -1, -10, \ldots)$, which does not
satisfy condition (1), but blowing down the ($-1$)-curves in this chain
gives $\chi_8$.

It is fairly straightforward to check that any other periodic chain
that satisfies condition (1) but that does not satisfy the maximality
condition (2) will become disconnected when blown down and will not
satisfy condition (3).  For example, consider replacing the 
($-6$)-curves in $\chi_6$ with ($-5$)-curves.  This does not satisfy condition (2).
Blowing down all ($-1$)-curves in this chain would give the chain $(-1,
-3, -1, -3, \ldots)$; blowing down again gives the chain $(-1, -1, -1,
\ldots)$, which becomes disconnected when any further ($-1$)-curves are
blown down.  Thus, conditions (1) and (2) together are apparently
necessary and sufficient to ensure that condition (3) holds.

Just as for the chain (\ref{eq:chain-4}), we can use the bounds on the
number of tensors computed above to estimate the maximum length
possible for the chains  $\chi_n$.  For $\chi_6$, the gauge algebra is $N(\ge_6
\oplus \gsu(3))$ or $({N \pm 1})\ge_6 \oplus N(\gsu(3))$, depending upon
which algebras appear on the two ends of the chain.  From
Table~\ref{f:primitives}, the increase in the tensor bound $\Delta T$
is 8/3 for each $\ge_6$ summand and $1/3$ for each $\gsu(3)$ summand.  If
the number of summands is equal to $N$ for each algebra, this gives
$\Delta T = 3 N$.  The number of blow-downs necessary to bring each
chain segment $(-6, -1, -3, -1)$ to a single $(-2)$ is $3$.  So as in
the example (\ref{eq:chain-4}), the number of tensors needed is of order $T
\sim{\cal O} (4 N)$, and we have $T \sim 4 N \leq 3 N + 9$, so again
the maximum number of cycles in the periodic chain will be of the
order of $N \sim{\cal O} (9)$.  A very similar analysis holds for
chain $\chi_8$, where each gauge algebra  component $\ge_7\oplus (\gsu(2) \oplus
\gso(7) \oplus \gsu(2))$ contributes $\Delta T = 5$ and requires $5$
blow-downs to bring each segment to the form $(-2)$.

The final chain type, $\chi_{12}$, is perhaps most interesting.  This is the
type of chain that appears in the F-theory realization of the model
with largest known $T$ \cite{Aspinwall-Morrison-instantons}.  Each
segment in the chain gives a contribution to the algebra of $\ge_8
\oplus \gf_4 \oplus 2(\gg_2 \oplus \gsu(2))$, and gives an increase
in $\Delta T$ of $25/3 + 16/9 + 5/6 = 10 \frac{17}{18}$, while
requiring 11 blow-downs to get to a single $-2$ from each segment in
the chain.  Since $\ge_8$ contributes the greatest part of the
increase to the bound on $T$, to maximize the size of the chain it is
desirable to use boundary conditions where an $\ge_8$ terminates the
chain on both sides.  With $N$ copies of the basic link and this
boundary condition, the gauge algebra is
\begin{equation}
\G = ({N + 1})\ge_8 \oplus N (\gf_4) \oplus 2 N (\gg_2 \oplus \gsu(2)) \,.
% \label{eq:}
\end{equation}
The upper bound on the number of tensors possible for this gauge algebra
is
\begin{equation}
T \leq 9 +  N \times \frac{197}{18}  + \frac{25}{3}    \,,
\label{eq:e8-bound}
\end{equation}
while the number of blow-downs needed to reduce to a single curve of
negative self-intersection is of the order of $12 \times N$ (one for
each curve removed).
Estimating
\begin{equation}
12 N_{\rm max} \cong \frac{52}{3}  + \frac{197}{18}  N_{\rm max} \; \; \Rightarrow
\; \; \frac{19}{18}  N_{\rm max} \cong \frac{52}{3} 
\; \; \Rightarrow \; \;
N < N_{\rm max} \cong 16.4 \,.
% \label{eq:}
\end{equation}
In fact, this configuration can be realized for $N = 16$, with $T =
193$.  As in the case of the chain (\ref{eq:chain-4}), this can be
done by attaching an extra pair of ($-1$)-curves to the next-to-last
$\ge_8$ summands in the chain.  We can check that a sequence of this type
with $N = 17$ is not possible.  Plugging $N = 17$ into
(\ref{eq:e8-bound}) gives $T \leq 203.4$.  But to remove all but one
negative self-intersection divisors from the chain would require
$17*12 = 204$ tensors, so we would need at least $T = 205$ even
without adding additional ($-1$)-curves as are needed in the $N = 16$
case.  Thus, the bounds we have determined here give a clear limit to
the size of sequences of this type.  As mentioned above, a
configuration of this type with $N = 16$ was identified in
\cite{Aspinwall-Morrison-instantons} and represents the model with the
largest known gauge group and value of $T$.  In a sequel to this paper
\cite{mt-toric} we show that this model naturally fits into the
framework of toric F-theory bases, and is indeed the maximal value of
$T$ that can be attained in that context.

\section{Conclusions}
\label{sec:conclusions}

We have shown that certain configurations of intersecting divisors on
the base surface of a 6D F-theory model give rise to gauge algebras and
matter content that cannot be removed by Higgsing scalar fields in the
theory.  We have tabulated all such possible ``non-Higgsable clusters''
(NHC's) in Table~\ref{f:primitives}.  Any base of a consistent 6D
F-theory model contains some number of NHC's connected by ($-1$)-curves
that do not contribute to the nonabelian gauge algebra.  As for generalized
del Pezzo surfaces, additional combinations of ($-2$)-curves that do not
connect to the NHC's can appear in special limits of the moduli but
do not affect the maximally Higgsed gauge algebra or matter
content.

We have also determined a bound on the number of tensor fields $T$ in
a theory based on the NHC content of the theory.  For models with no
nonabelian gauge algebra in the maximally Higgsed phase, $T \leq 9$.
Each NHC contributes a fixed positive quantity to the upper bound on
$T$ as described in Table~\ref{f:primitives}.  Taken together, the
classification of NHC's and the upper bound on $T$ bound and
characterize the bases possible for 6D F-theory models.

While we have identified the geometry of all divisor combinations
giving NHC's, there are several things that we have not proven.
First, while the gauge and matter content associated with each NHC
cannot be Higgsed, we have not proven that every F-theory model has
sufficient degrees of freedom to Higgs all fields down to the minimal
NHC content.  While we believe that this is true, it is possible in
principle that there some complicated F-theory models might not have
enough scalar degrees of freedom to break the gauge algebra and matter
content through Higgsing to the minimal spectrum required by the NHC's.  Second, our analysis has been carried out purely in the
context of F-theory.  The question of possible gauge groups (or algebras)
and matter
content in the maximally Higgsed phase can be posed more generally for
the class of all quantum-consistent 6D supergravity models.  
There are 6D supergravity models satisfying all known
quantum consistency constraints that cannot be realized in F-theory
\cite{KMT, KMT-II, 0}.  For example, there is an apparently-consistent
6D supergravity model containing an $\gsu(8)$ gauge algebra and matter in
the ``box'' ({\bf 336}) representation that violates the Kodaira
constraint from F-theory.  It is possible that there are 
``non-Higgsable'' gauge algebra and matter factors that cannot appear in
F-theory but that are possible in consistent 6D supergravity models.
We conjecture that the NHC content identified in this paper describes
all maximally Higgsed 6D supergravity theories, whether realized
through F-theory or not.  Further analysis of this question is left as
an open problem for future work.

By combining the NHC structure identified in this paper with bounds on
the number of tensors and the geometry of blow-up processes, it should
be possible in principle to systematically identify all bases for 6D
F-theory models.  In a sequel to this paper \cite{mt-toric} we
carry out such an analysis in the case of toric bases.  For each
F-theory base, a wide range of different gauge groups and matter
content can be realized by allowing the elliptic fibration to become
more singular without requiring a blow-up of the base --- this
corresponds to ``un-Higgsing'' the theory away
from the maximally Higgsed content determined from the NHC structure
on the base.  Recent papers describe some of the range of
possibilities of models that can be realized over simple bases such as
$\P^2$ and $\F_m$ \cite{KMT, 0, Braun, Morrison-Taylor}.  Combining
analysis of bases through the methods of this paper with such analyses
for each base gives tools for understanding the complete space of 6D
F-theory models.

While the results of this paper are directly relevant only for
six-dimensional theories, the general philosophy and ideas underlying
this analysis should have an analogue in four dimensions, where the
space of theories is much richer and more complex.  In four dimensions
there are a number of additional subtleties and issues that would need
to be addressed for any kind of systematic analysis.  In particular,
the intersection form is a triple intersection product, and is less
well understood than the intersection form on complex surfaces
relevant in the 6D case.  Also, in four dimensions, fluxes
complicate the story by lifting moduli and modifying the set of
massless degrees of freedom.  Nonetheless, it may be possible in four
dimensions to identify some analogue of the ``non-Higgsable cluster''
structures we have found in 6D that characterize many key aspects of
the theories.  We leave investigation of these questions in four
dimensions to future work.

\vspace*{0.1in}

{\bf Acknowledgements}: We would like to thank Antonella Grassi,
Thomas Grimm, Vijay
Kumar, Joe Marsano, and
Daniel Park for helpful discussions.  Thanks to the the Aspen
Center for Physics for hospitality while this work was carried out.
This research was supported by the DOE under contract
\#DE-FC02-94ER40818, and by the National Science Foundation under
grant DMS-1007414

\newpage

\appendix

\section{Appendix: systematic analysis of $(-1)$-curves intersecting NHC's}

In this Appendix we summarize the results of a systematic analysis of which
combinations of NHC's can be intersected by a single ($-1$)-curve on a
valid F-theory base.  The table below gives a list of all possible
ways in which a ($-1$)-curve can intersect a single NHC.  The NHC is
denoted by an ordered $k$-tuple of self-intersection numbers for the
irreducible effective curves comprising the NHC.  The number of times
the ($-1$)-curve intersects each curve in the
NHC is denoted by the
number of dots over the number representing each curve.  The possible
intersections of a ($-1$)-curve with a single NHC are indexed by an
integer $n$.  For each $n$ the possible pairs $[n, n']$ of NHC
intersections by a ($-1$)-curve are listed (listing only values $n' \geq
n$; the set of possible pairs is symmetric under interchange $n
\leftrightarrow n$).  Following each $n'$ denoting a pair, a list is
given of possible values $n''\geq n'$ for which the triplet $[n, n',
  n'']$ of intersections is allowed.  (For example, a single 
($-4$)-curve can appear in the triplet combinations $[6, 10, 16], [6, 16,
  16],$ and $[6, 16, 25]$.)
Thus, the table contains all
possible combinations of 1, 2, or 3 NHC's that can be intersected by a
($-1$)-curve.  There are also 6 combinations of 4 NHC intersections
possible.  These are given by all configurations containing  2
intersections of type 16 (a $(-3, -2, -2)$ NHC intersected along the
last $-2$), and another two intersections that are any combination of
types 10, 16, and 25.  ({\it i.e.}, the 6 possible configurations
connecting 4 NHC's are
[10, 10, 16, 16], [10, 16, 16,
  16], [10, 16, 16, 25], [16, 16, 16, 16], [16, 16, 16, 25], [16, 16,
  25, 25]).  This gives a total of 183 possible ways in which a single
($-1$)-curve can intersect a combination of 1, 2, 3, or 4 NHC's.  As
discussed in the main text, the gauge group and matter content  of
these configurations are all determined by the product of the factors
associated with each separate NHC.

A few comments may be helpful regarding the analysis for anyone
interested in reproducing these or similar computations.
To determine the degree of vanishing of $-nK$ on any system of curves
$\{C_i\}$, we must identify the minimal set of non-negative integral
values $c_i$ so that
\begin{equation}
-nK = \sum_{i}c_i  C_i+ X \,,
\label{eq:k-equation}
\end{equation}
where $X \cdot C_j \geq 0 \;\forall j$.  For some systems there is no
set of values $c_i$ that satisfy these conditions.
We have used the following algorithm to determine the solution for
$c_i$: First, we determine the minimum values $c_i^{(0)}$ needed for
each curve independently ({\it i.e.}, ignoring intersections between
distinct curves).  We then iteratively increase the values $c_i$ to
compensate for intersections using the previous values
at each stage.  Defining $M_{ij} = C_i \cdot C_j$, we thus have
\begin{equation}
c_i^{(0)} =\lceil n (m-2)/m \rceil,
% \label{eq:}
\end{equation}
just as in the case of a single divisor discussed in Section
\ref{sec:general}.
Defining $v_i = -nK \cdot C_i = n (2-m)$, we then use the difference between the
desired value on the LHS and the value computed using $c_i^{(0)}$ to
determine the next correction (which is not allowed to be negative)
\begin{equation}
c_i^{(1)} = c_i^{(0)} + {\rm Max} (0, \lceil
\frac{v_i-M_{ij}c_j^{(0)}}{-m}\rceil) \,,
% \label{eq:}
\end{equation}
and so forth.  When this process converges we can test to check if any
intersection points have vanishing degrees 4, 6, 12 or above
indicating a singularity; when the process does not converge the
values continue to climb, indicating the absence of a solution.  This
method was used to determine all valid intersecting curve combinations
described in this Appendix.

The solution to the degree equation (\ref{eq:k-equation}) also can be
characterized in terms of linear algebraic inequalities.
Stating the conditions on the $c_i$'s in terms of inequalities, we have the
conditions
\begin{equation}
c_i = (M^{-1})_{ij} (v_j -z_j)
\label{eq:c-equations}
\end{equation}
where $z_j \geq 0$ are non-negative integers.  For any given system we can
compute $M^{-1}$.  This then defines a cone of values within $\Q^n$
containing the points defined by the RHS of (\ref{eq:c-equations}) for
different $n$-tuples of non-negative integers $z_i$; the solution
$c_i$ must lie in the intersection between this cone and the sector of
non-negative integer $n$-tuples $c_i$.
This characterization can be used to identify situations where there
is no solution, but the analysis can be slightly subtle.  For example,
in the case of a ($-3, -2, -2$) NHC intersecting a ($-1$)-curve along
each of the 3 curves in the NHC, the linear algebra gives a rational
value for $M^{-1} v$ that is positive in all components, but there is
no vector of non-negative  integers $z_j$ such that the resulting
$c_i$ is non-negative itself for $-4K$, while a solution exists for
$-6K$ and $-12K$.  This approach also does not uniquely
determine the solution for $c_i$.  For example, consider a $-1$
connecting the NHC's $(-3, -2, -\dot{2})$, $(-\dot{3}), (-\dot{3})$.
Working out the corresponding intersection matrix shows that
(\ref{eq:c-equations}) gives a solution with all $c_i = n/2$ when $z_j
= 0$ for all $j$.  This is, however, a singular configuration, and is
not the minimal solution possible.  The algorithm described in the
preceding paragraph
gives the correct solution in this case, which has vanishing degree 0
for $g$ and $\Delta$ along the ($-1$)-curve and no change in the degrees
of vanishing of these multiples of $-K$ on the remaining curves, as
can immediately be seen from the fact that condition (\ref{eq:cp}) is
satisfied for $n = 6, 12$.

\begin{table}
\begin{center}
\begin{tabular}{|r | r |l  |}
\hline
Index ($n$) &
NHC &
pairs \{triplets\} \\ \hline \hline
1 & $(-\dot{12})$ & 16 \\
2 & $(-\dot{8})$ & 10, 16, 25\\
3 & $(-\dot{7})$ & 10, 16, 25\\
4 & $(-\dot{6})$ & 7, 10, 15, 16, 25\\
5 & $(-\dot{5})$ & 7, 9, 10, 14, 15, 16, 25\\
6 & $(-\dot{4})$ & 6, 7, 9, 10 \{16\}, 14, 15, 16 \{16, 25\}, 22, 25, 26\\
7 & $(-\dot{3})$ & 7 \{7, 10, 16\}, 9, 10 \{10, 16, 25\}, 13, 14, 15,
    16  \{16, 25\}, \\
& & \hspace*{0.1in}22, 25 \{25\}, 26, 27, 29\\
8 & $(-\ddot{3})$ & \\
9 & $(-\dot{3}, -{2})$ & 9, 10 \{10, 16\}, 14, 15, 16 \{16, 25\}, 22, 25, 26\\
10 & $(-{3}, -\dot{2})$ & 10 \{10, 14, 15, 16, 22, 25\}, 12, 13, 14 \{16\},
    15 \{16, 25\}, \\
& & \hspace*{0.1in}16 \{16, 22, 25\}, 
19, 20, 22 \{25\}, 25 \{25\}, 26, 27,
   29\\
11 & $(-\ddot{3}, -{2})$ & \\
12 & $(-\dot{3}, -\dot{2})$ & 16\\
13 & $(-{3}, -\ddot{2})$ & 15, 16 \{16\}, 22, 25\\
14 & $(-\dot{3}, -{2}, -{2})$ & 14, 15, 16 \{16, 25\}, 22, 25, 26\\
15 & $(-{3}, -\dot{2}, -{2})$ & 15, 16 \{16, 25\}, 22, 25 \{25\}, 26, 27, 29\\
16 & $(-{3}, -{2}, -\dot{2})$ & 
 16 \{16, 22, 25, 26, 27, 29\}, 19, 20, 22 \{25\}, 25 \{25\}, \\
& & \hspace*{0.1in}26, 27, 29\\
17 & $(-\ddot{3}, -{2}, -{2})$ & \\
18 & $(-\dot{3}, -\dot{2}, -{2})$ & \\
19 & $(-\dot{3}, -{2}, -\dot{2})$ & 25\\
20 & $(-{3}, -\dot{2}, -\dot{2})$ & 25\\
21 & $(-{3}, -\ddot{2}, -{2})$ & \\
22 & $(-{3}, -{2}, -\ddot{2})$ & 22, 25 \{25\}, 26, 27, 29\\
23 & $(-\dot{3}, -{2}, -\ddot{2})$ & \\
24 & $(-{3}, -\dot{2}, -\ddot{2})$ & \\
25 & $(-\dot{2}, -{3}, -{2})$ & 25 \{25\}, 26, 27, 29\\
26 & $(-{2}, -\dot{3}, -{2})$ & 26\\
27 & $(-\ddot{2}, -{3}, -{2})$ & \\
28 & $(-\dot{2}, -\dot{3}, -{2})$ & \\
29 & $(-\dot{2}, -{3}, -\dot{2})$ & \\
30 & $(-\dot{2}, -{3}, -\ddot{2})$ & \\
31 & $(-\ddot{4})$ & \\
\hline
\end{tabular}
\end{center}
\caption[x]{\footnotesize Table of all ways in which a ($-1$)-curve can
  intersect a single non-Higgsable cluster of irreducible effective
  divisors carrying a nonabelian gauge group.  Each cluster (NHC) is
  denoted by  a list of self-intersections of curves connected in a
  linear chain, with dots
  indicating intersection with the ($-1$)-curve.  The last column of the
table includes
  information about all pairs and triplets of NHC's that can be
  intersected by the ($-1$)-curve, as described in the main text of
  the Appendix.}
\label{t:intersections}
\end{table}

\end{document}